\documentclass[pra,preprint]{revtex4-1}
\usepackage{hyperref} 

\usepackage{graphicx} 
\usepackage{subfig}
\usepackage{color}
\usepackage{filecontents}
\usepackage{soul}
\usepackage{physics}
\usepackage{mathrsfs}

\usepackage{relsize}
\usepackage{amsmath}
\usepackage{xcolor}
\usepackage{color, colortbl}
\definecolor{LightCyan}{rgb}{0.88,1,1}
\definecolor{corn}{rgb}{0.98, 0.93, 0.36}
\definecolor{pastelyellow}{rgb}{0.99, 0.99, 0.59}
\usepackage{tabularx,stackengine}

\newcommand{\be}{\begin{equation}}
\newcommand{\ee}{\end{equation}}
\newcommand{\bea}{\begin{eqnarray}}
\newcommand{\eea}{\end{eqnarray}}
\newcommand*{\myeqref}[2][Eq.~]{%
  \hyperref[{#2}]{#1(\ref*{#2})}%
}
\def\equationautorefname#1#2\null{%
  Eq.#1(#2\null)%
}

\usepackage{dcolumn}
\usepackage{bm}
\usepackage{amssymb,amsmath}
\usepackage{color}
\usepackage{float}
\usepackage{tikz}
\usetikzlibrary{arrows}

\definecolor{DarkGreen}{rgb}{0,0.6,0.2}

\usepackage{accents}

\begin{document}
\title{Non-Markovianity in photosynthetic reaction centers: A noise-induced quantum coherence perspective}
\author{Zibo Wang$^{1,2}$, Antonio V. Lim$^{1}$, and Imran M. Mirza$^{1,\ast}$}
\affiliation{$^{1}$Macklin Quantum Information Sciences, Department of Physics, Miami University, Oxford, Ohio 45056, USA\\
$^{2}$Department of Physics, Applied Physics and Astronomy, Rensselaer Polytechnic Institute, Troy, NY 12180, USA}
\email{mirzaim@miamioh.edu}

\begin{abstract}
The long-standing problem of nearly perfect photosynthetic yield in some types of bacteria and nearly all kinds of plants despite the interaction with a hot and noisy environment has witnessed quantum optical explanations in the last decade. Typically in these explanations, photosynthetic reaction centers are modeled as five-level quantum heat engines where the generation of Fano-type interference due to the coupling of discrete state transitions with a common Markovian reservoir is held responsible for the enhancement of the photosynthetic efficiency. In this work, we go beyond the Born-Markov approximation used in the earlier works and study the impact of non-Markovian environments with Lorentzian spectral densities on the dynamics of light-harvesting complexes.
\end{abstract}

\maketitle

\section{Introduction}
The origins of the conceptual development of quantum biology can be traced back to the early days of quantum mechanics \cite{schrodinger1992life}. However, it is only in the recent decades, studies in chemistry, biology, and physics have theoretically proposed and experimentally indicated that specific types of biological processes may be taking advantage of quantum effects to achieve high levels of performance \cite{mohseni2014quantum}. Some of the captivating examples in this regard include magnetoreception-based navigation in some species of birds (for example, homing pigeons) \cite{gauger2011sustained,walcott1996pigeon}, long-range quantum tunneling of electrons in protein complexes/redox chains \cite{gray2003electron,winkler2014long}, Dyson and Wright quantum theory of olfaction based on molecular vibrations \cite{malcolm1938scientific,wright1977odor}, and the presence of quantum coherence in photosynthesis enabling higher yield in light-harvesting complexes \cite{engel2007evidence,ishizaki2009theoretical}, etc. 

\begin{figure}
\centering
\includegraphics[width=2.25in, height=1.9in]{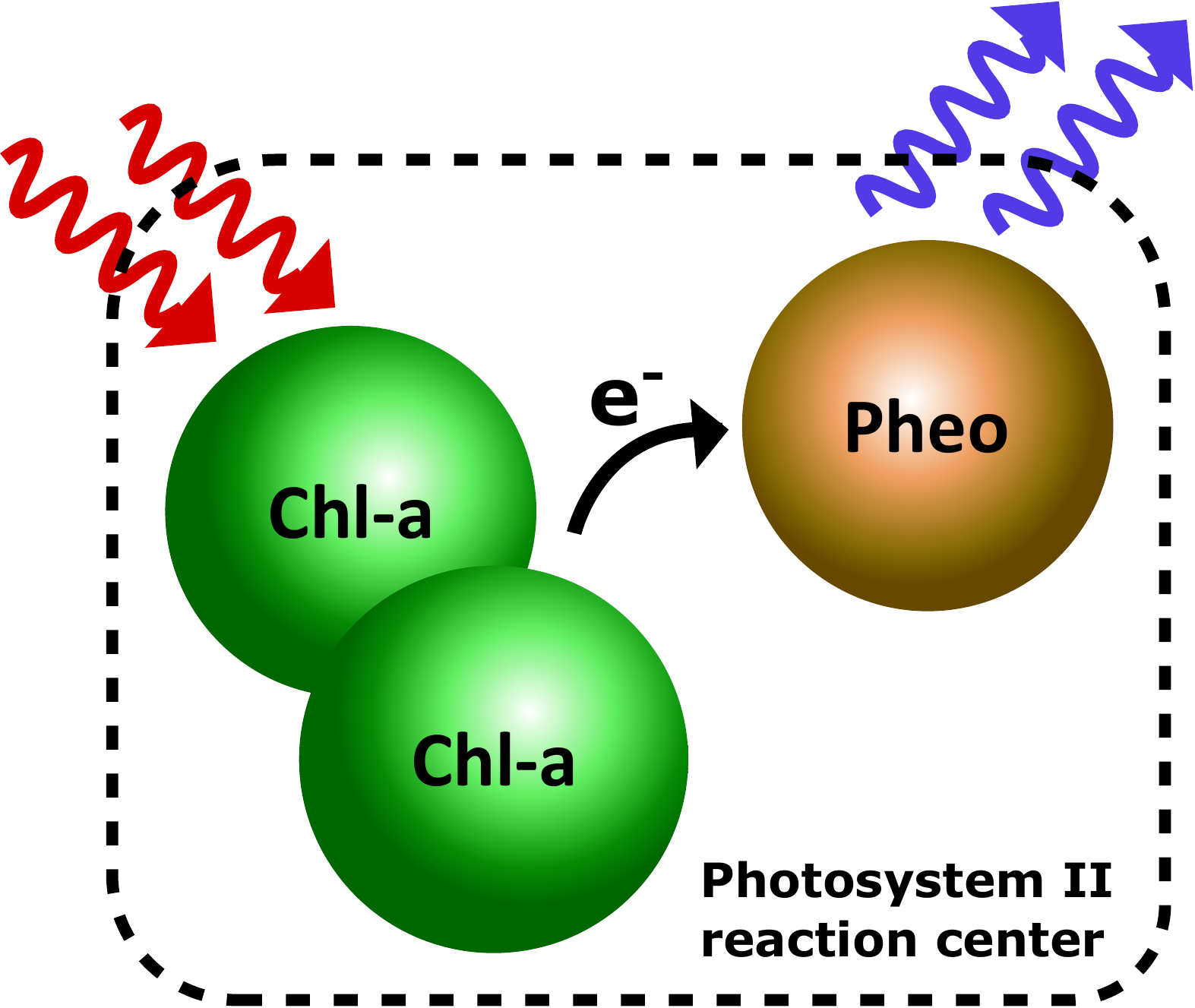} 
\captionsetup{
  format=plain,
  margin=1em,
  justification=raggedright,
  singlelinecheck=false
}
\caption{Schematic representation of a photosynthetic reaction center in type-II photosystems.}
\label{Fig1}
\end{figure}

In the present work, we focus on the intriguing problem of photosynthesis-based light-harvesting complexes (protein-based Fenna–Matthews–Olson protein as a preliminary example). There in the so-called photosystem Type-II reaction centers \cite{michel1988relevance} two Chlorophyll molecules (Chl-a) absorb solar photons while acting as the electron donor. Followed by that the photons are transmitted to a Pheophytin (Pheo) molecule which act as electron acceptors, as shown in Fig.~\ref{Fig1}. As a result of this photon exchange, charge separation is generated in this tri-pigment setup which leads to the conversion of sunlight energy into chemical energy. In the last decade or so, quantum optical models have been reported to mimic the process of sunlight-to-chemical energy conversion \cite{cheng2009dynamics,cao2020quantum} in photosystems. In these models, the coupling between the photosynthetic reaction centers with their immediate surroundings (environment) is modeled as Markovian in nature essentially meaning that the environments fails to keep any record/memory of its coupling with the reaction centers.
As an example, Dorfman et al. studied this problem under the Markov and Born approximations (the Born approximation assumes a weak type of system-bath interaction) and discussed how a typical photosynthesis process can be modeled as a five-level quantum heat engine (QHE) dissipatively coupled to high and low temperate reservoirs \cite{dorfman2013photosynthetic, wang2020dissipative}. They showed that the coupling of certain discrete energy transitions (in the five-level system) with a common Markovian environment generates noise-induced coherence in the system that can aid increasing photosynthetic yield by 27\% compared with no coherence case. However, in reality (see for instance Ref.~\cite{lee2007coherence}) the interaction of the reaction centers with the surrounding protein environment can be rather strong which severely challenges the validity of the Markov approximation. Therefore, to perform a more realistic treatment of the problem, in the present work we go beyond the Markov approximation and study the dynamics of photosynthetic complexes under memory-full/non-Markovian conditions.

In particular, we derive the time convolutionless (TCL) type non-Markovian master equation \cite{breuer2002theory} for five-level quantum systems and apply it to analyze the light absorption and exciton transfer in photosynthesis. As one of the key results, we find that non-Markovian environments with a Lorentzian spectral density can generate and (more importantly) control the noise-induced coherence in the system. This opens the possibility of increasing the efficiency of the photosynthesis process in a transient manner which is not possible to achieve with traditional Markovian environments.

\section{Theoretical Description}
\begin{figure}
\centering
\fbox{\includegraphics[width=4.85in, height=1.8in]{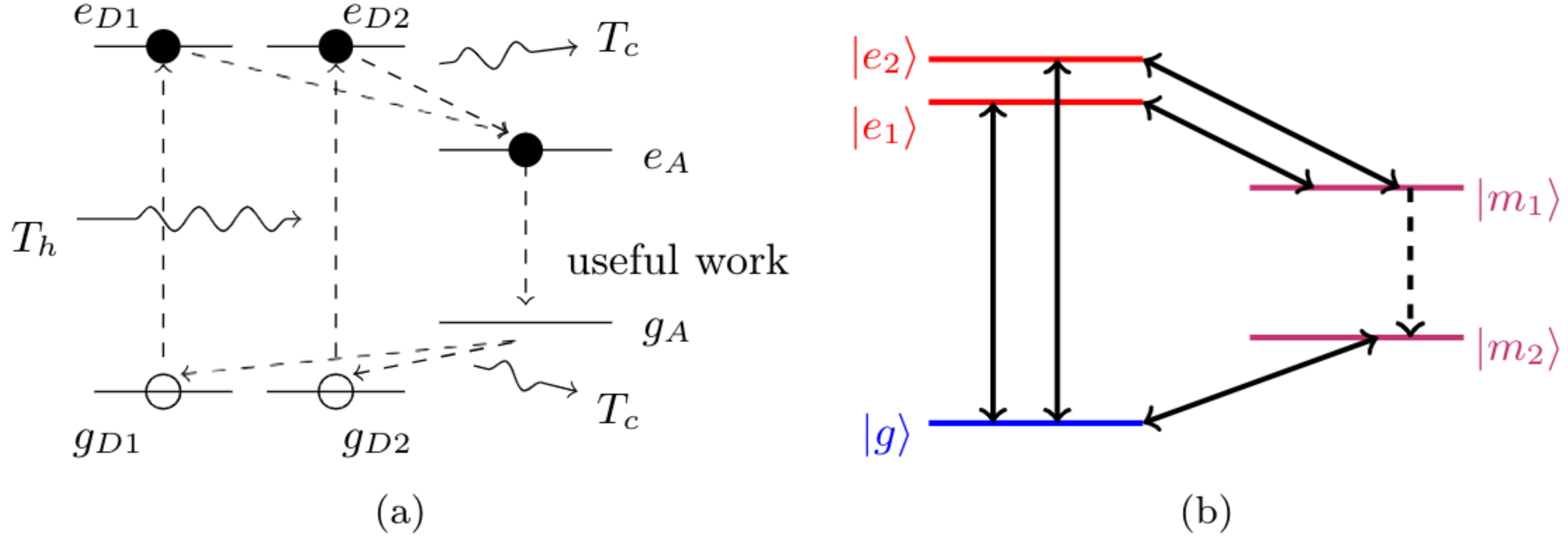}}
\captionsetup{
  format=plain,
  margin=1em,
  justification=raggedright,
  singlelinecheck=false
}
\caption{Energy-level configuration for (a) the acceptor-donors complex, and (b) the photosynthesis reaction center in its entirety.}
\label{Fig2}
\end{figure}

In Fig.~\ref{Fig2}(a) we represent the energy-level diagram for our five-level donor-acceptor complex. The thermodynamics cycle begins with the absorption of thermal photons from the high-temperature reservoir at temperature $T_h$ with the excitation of electrons in the donor levels. Upon de-excitation, the electrons transfer to the acceptor molecules with emission of photons into a heat sink with temperature $T_c$. Useful work is extracted out between the ground state $g_A$ and excited states $e_A$ of the acceptor molecules. In Fig.~\ref{Fig2}(b) we draw the energy-level diagram at the reaction center level. Therein, the state $\ket{g}$ specifies the ground state for the entire molecular complex. The state $\ket{e}$ (split into two levels $\ket{e_1}$ and $\ket{e_2}$) shows the single donor molecule excited state. $\ket{m_1}$ is the meta-stable state with the charge being separated and transferred to the acceptor molecul followed by the creation of a hole in the donor molecule. The other meta-stable energy state $\ket{m_2}$ describes the ionized state with the photon supplied to the sink and the useful work being performed. Finally, the system undergoes the transition $\ket{m_2}\rightarrow\ket{g}$ through the spontaneous emission.

\section{Non-Markovian Master Equation for Damped Few-Level Atoms}
Next, to describe the non-Markovian dynamics of the reaction centers we apply the master equations technique. Typically, a projection operator framework is adopted which leads to two types of equations namely the Nakajima-Zwanzig master equation (which has an integrodifferential form) \cite{nakajima1958quantum, zwanzig1960ensemble} and the time-convolutionless (TCL) master equation (which is a local in time differential equation) \cite{shibata1980expansion}. Keeping in view the need for a numerical solution for our five-level QHE problem,  we focus on the TCL master equation which takes the following form for the case of a driven dissipative two-level system \cite{haikka2010non} (here $\hbar=1$):
\begin{align}
&\frac{d\hat{\rho}(t)}{dt}=-i[\hat{H}_S+\hat{H}_L,\hat{\rho}]+\sum_{\lbrace s=+,-\rbrace}\Big[C^2_s\gamma(t)\Big(\hat{S}_{-s}\hat{\rho}\hat{S}_s-\frac{1}{2}\lbrace\hat{S}_s\hat{S}_{-s},\hat{\rho}\rbrace\Big)\nonumber+sC_sC_{0}\gamma(t)\Big(\hat{S}_{-s}\hat{\rho}\hat{S}_z\\
&+\hat{S}_z\hat{\rho}\hat{S}_s\Big)\Big]+C^2_0\gamma(t)[\hat{S}_z\hat{\rho}\hat{S}_z-\frac{1}{2}\lbrace \hat{S}_z\hat{S}_z,\hat{\rho}\rbrace]+C^2_0\gamma(t)[\hat{S}_+\hat{\rho}\hat{S}_++\hat{S}_+\hat{\rho}\hat{S}_-]+C_0[\frac{\gamma(t)}{2}\lbrace \hat{S}_x,\hat{\rho}\rbrace\nonumber\\
&+i\lambda\lbrace \hat{S}_x,\hat{\rho} \rbrace],\label{NMME2}
\end{align}
where $C_{\pm}=\frac{\omega_{eg}+\Delta}{2\omega_{eg}}$, and $C_0=\frac{\Omega}{2\omega_{eg}}$ with $\omega_{eg}$, $\Delta$, and $\Omega$ being the atomic transition frequency, laser detuning and the Rabi frequency, respectively. $\hat{H}_S$ is the system Hamiltonian while $\hat{H}_L$ represents the Lamb Hamiltonian given by 
\begin{align}
\hat{H}_L=\lambda(t)[C^2_+\hat{\sigma}_-\hat{\sigma}_++C^2_+\hat{\sigma}_-\hat{\sigma}_-+C^2_0\hat{\sigma}^2_z].
\end{align}
In the last two equations the time-dependent parameters $\gamma(t)$ and $\lambda(t)$ which have the interpretation of atomic decay rate and Lamb shift are related to the transition rate $\Gamma(t)$ through $\Gamma(t)=\gamma(t)/2+i\lambda(t)$, while $\Gamma(t)$ itself is given by
\begin{equation}
\Gamma(t)=\int_0^tdt'\int d\omega'J(\omega')\exp[i(\omega_L-\omega')t'].
\end{equation}
Note the subtle but important point that in the Markovian Lindblad master equations for two-level quantum systems both atomic decay rate and Lamb shifts were time-independent \cite{breuer2002theory}.
For our photosynthesis model, coherence is not laser-assisted; rather the coupling of the excited donor states with a common ground state indirectly produces noise-induced coherence through a Fano-like process \cite{dorfman2013photosynthetic}. Therefore, we ignore the driving laser, and after some mathematically involved calculations (for details see our thesis \cite{wang2021quantum}), we extend the aforementioned theory to a five-level situation in the interaction picture. Finally, we find the following TCL non-Markovian master equation 
\begin{align}
\frac{d\tilde{\rho}(t)}{dt}&=\sum_{i,j=1}^5\gamma_{ij}(t)\bigg[(\bar{n}_{i}+1)\Big(\hat{S}_i^\dagger \hat{S}_j\tilde{\rho}(t)+\tilde{\rho}\hat{S}_i^\dagger \hat{S}_j-\hat{S}_j\tilde{\rho}(t)\hat{S}_j^\dagger-\hat{S}_j\tilde{\rho} \hat{S}_i^\dagger\Big)\nonumber\\
&+\bar{n}_{i}\Big(\hat{S}_j\hat{S}_j^\dagger\tilde{\rho}+\tilde{\rho}(t) \hat{S}_j\hat{S}_j^\dagger-\hat{S}_i^\dagger\tilde{\rho}\hat{S}_j-\hat{S}_j^\dagger\tilde{\rho}\hat{S}_j\Big)\bigg],\label{NMME5}
\end{align}
where we have ignored the Lamb shift. Note that, unlike the two-level atom case, we have now included the non-zero temperature of the baths (represented through the Bosonic occupation number $\bar{n}_i$) to incorporate the thermal aspect of the photosynthesis process. The time-dependent transition rate for the present case takes the form
\begin{equation}
\gamma_{ij}(t)=\int_0^t dt'\int d\omega'J(\omega')\mathrm{e}^{i(\omega_{ij}-\omega')(t-t')}.\label{gammaT}
\end{equation}
As pointed out above, the transition rate in the non-Markovian case becomes time-dependent and varies as a function of the environmental spectral density function $J(\omega)$. This, in complete contrast to the Markovian case, leads to a more realistic situation where the environment now keeps a record of the memory effects. Proceeding further, we calculate the time-dependent transition rate $\gamma_{ij}(t)$ and, for that, we assume the following Lorentzian form of the density function
\begin{equation}
J(\omega)=\mathcal{N}^{-1}\pi\gamma[1+\gamma^{-1}(\omega-\omega_0)^2],
\end{equation}
where $\omega_0$ represents the frequency at which the density function's peak resides, $\gamma$ shows the width and, $\mathcal{N}$ is the normalization constant. We remark that this form of the spectral density is experimentally feasible and can be mimicked by a single-mode optical cavity with imperfect mirrors \cite{raizen1989normal}. To perform the time integral in Eq.~\eqref{gammaT} we consider the following complex function $f(z)$ which we assume to be holomorphic on a contour $\mathrm{C}$
\begin{equation}
f(z)=\frac{\mathrm{e}^{-2\pi i\gamma z}}{z^2+t^2}.
\end{equation}
To integrate $f(z)$ we focus on three possible cases
\begin{figure}[t]
\begin{center}
\begin{tikzpicture}
\draw [->, line width=1.2] (-2,0) -- (2.1,0) node [label={[shift={(0.25,-0.5)}]$\mathrm{R}$}]{};
\draw [->, line width=1.2] (0,-1) -- (0,2.1) node [label={[shift={(0.28,-0.46)}]$\mathrm{C}$}]{};
\filldraw[red] (0,0.5) circle (2pt) node [label={[shift={(0.25,-0.5)}]$it$}]{};
\filldraw[red] (0,-0.5) circle (2pt) node [label={[shift={(0.25,-0.5)}]$-it$}]{};
\draw [<-, blue, line width=1.1] (-1.8,0) arc (180:90:1.8);
\draw [blue, line width=1.1] (0,1.8) arc (90:0:1.8);
\draw [->, blue, line width=1.1] (-1.8,0) -- (0,0);
\draw [blue, line width=1.1] (0,0) -- (1.8,0);
\end{tikzpicture}
\end{center}
\vspace{-5mm}
\caption{A semi-circle contour for the integration of $f(z)$ on the
complex plane.}
\label{fig:complex}
\end{figure}
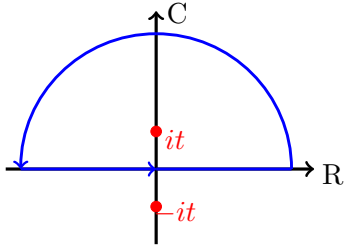
\begin{itemize}
\item {Case-(I) $\gamma=0$ (the trivial scenario):
\begin{align*}
\int_\mathrm{R}\frac{1}{x^2+t^2}=&\left.\frac{1}{t}\arctan(x)\right|_{x=-\infty}^{x=\infty}=\frac{\pi}{t}
\end{align*}
}
\item {Case-(II) $\gamma<0$: As shown in Fig.~\ref{fig:complex}, we consider a semicircle containing the pole at $it$. As the radius of the semi-circle grows and approaches infinity, we can ignore the arc contribution that leaves us with the integration carried out on the straight line. Consequently, Cauchy's residue theorem gives us
\begin{align*}
\int_\mathrm{R}f(z)dz = 2\pi i\left(\frac{1}{2it}\mathrm{e}^{2\pi\gamma t}\right)=\frac{\pi}{t}\mathrm{e}^{-2\pi|\gamma| t}.
\end{align*}}
\item {Case-(III) $\gamma>0$: Similar to the above $\gamma<0$ case leading to the same result. The main difference will now be to draw the contour below the $\mathrm{R}$-axis such that the pole point $-it$ can be enclosed by the contour.}
\end{itemize}
Putting everything together, we observe 
\begin{align}
&\frac{\mathcal{N}}{\pi\gamma}\int\frac{1}{1+\left(\frac{\omega'-\omega_0}{\gamma}\right)}\mathrm{e}^{i(\omega_0-\omega')(t-t')}d\omega'
=\mathcal{N}\mathrm{e}^{-\gamma(t-t')},
\end{align}
which leads to the following temporal integral
\begin{align}
\gamma(t)=\mathcal{N}\int_0^tdt'\mathrm{e}^{-\gamma(t-t')}
=\gamma(1-\mathrm{e}^{-\gamma t}).
\end{align}
We note that in the long-time limit i.e. as $t\rightarrow\infty$, the non-Markovian decay rate $\gamma(t)$ reduces to its Markovian counterpart $\gamma$. Extending this result to the time-dependent transition rate $\gamma_{ij}(t)$ between energy states $\ket{i}$ and $\ket{j}$ we can write
\begin{equation}
\gamma_{ij}(t)=\gamma_{ij}(1-\mathrm{e}^{-\gamma_{ij}t}),
\end{equation}
where (as before) $\gamma_{ij}$ is the Markovian decay rate for the transition $\ket{i}\longleftrightarrow\ket{j}$. Next, in Table \ref{paratable} we summarize the set of parameters used in three possible regimes of an open quantum system treatment of photosynthesis. We refer to systems as being in an overdamped (underdamped) regime if the energy separation between the states $\ket{e_1}$ and $\ket{e_2}$ is smaller (larger) than these states' inverse lifetimes. In the case that the energy gap is comparable to the inverse lifetime, we call such a scenario intermediately damped. Furthermore, for simplicity, we assume the temperature of the cold reservoirs to be almost zero compared to the hot reservoir temperature; i.e., we set $\bar{n}_{1c}\approx 0$, $\bar{n}_{2c}\approx 0$, and $\bar{n}_{m_2c}\approx 0$. Note that the set of parameters described in Table \ref{paratable} is experimentally feasible for type II photosystems as reported in  Ref.~\cite{abramavicius2010energy}.
\begin{table}[htbp]
\centering
\caption{\bf Parameters used under different working regimes}
\begin{tabular}{c|c|c|c}
\hline\hline
\hspace{-2mm}Parameter/Regime & Over & Under & Intermediate\\
\hline\hline
$\Delta [cm^{-1}]$ & 120 & 600 & 720 \\
$\gamma_{e_1g} [cm^{-1}]$ & 0.005 & 0.005 & 0.005 \\
$\gamma_{e_2g} [cm^{-1}]$ & 0.0016 & 0.005 & 0.005 \\
$\gamma_{e_1m} [cm^{-1}]$ & 140 & 35 & 280  \\
$\gamma_{e_2m} [cm^{-1}]$ & 18 & 35 & 280  \\
$\gamma_{m_1m_2} [cm^{-1}]$ & $10^{-10}$ & $10^{-10}$ & $10^{-10}$\\
$\gamma_{m_2g} [cm^{-1}]$ & 200 & 50 & 300 \\
$\tau^{-1}_2 [cm^{-1}]$ & 41 & 41 & 41 \\
$\bar{n}_{1h}$ & 60,000 & 10,000 & 90,000\\
$\bar{n}_{2h}$ & 10,000 & 20,000 & 10,000 \\
\hline\hline
\end{tabular}
\label{paratable}
\end{table}
\section{Results in the over-, under-, and intermediate coupling regimes}
\begin{figure*}
\centering 
\includegraphics[width=2.05in, height=1.25in]{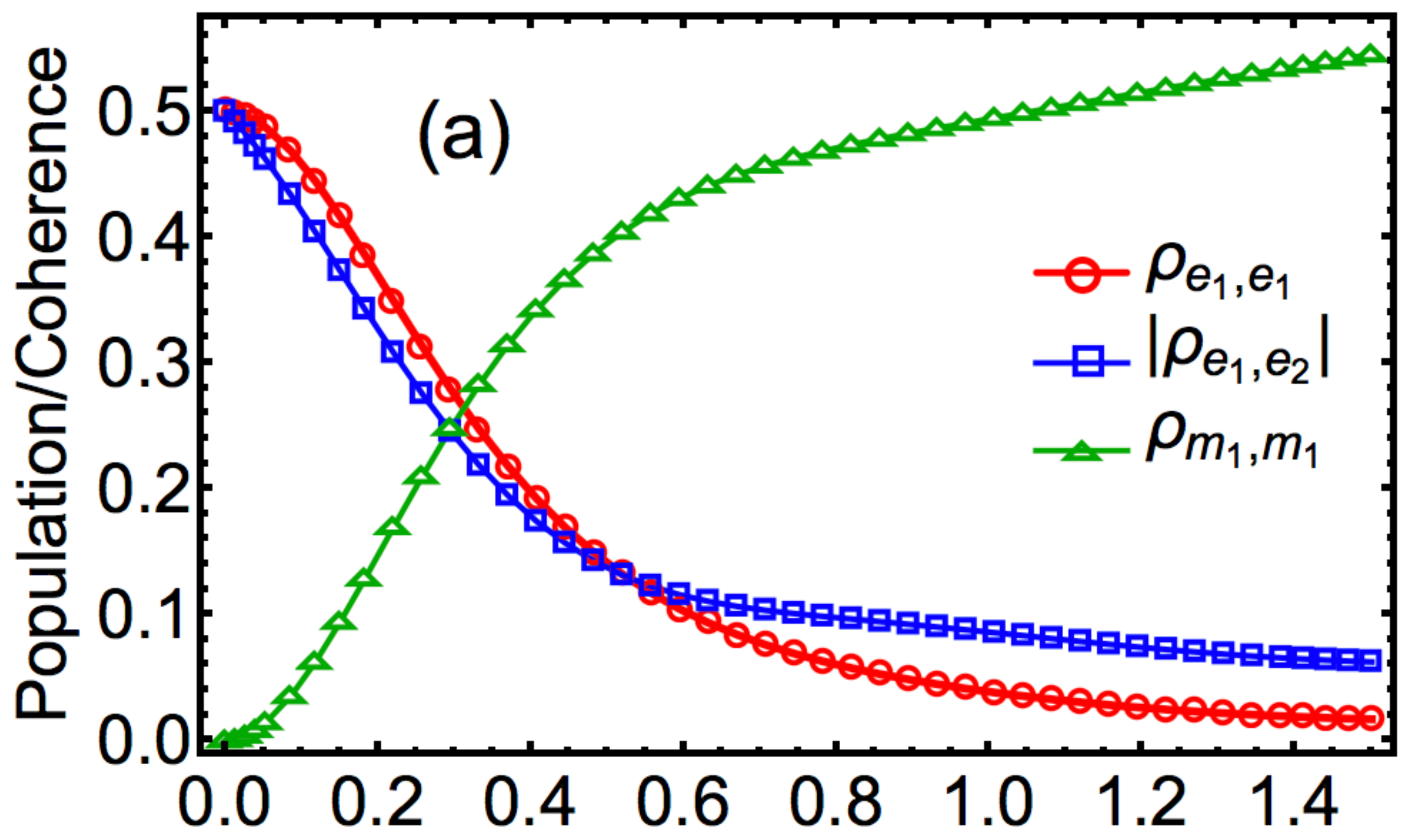}
\includegraphics[width=2.05in, height=1.25in]{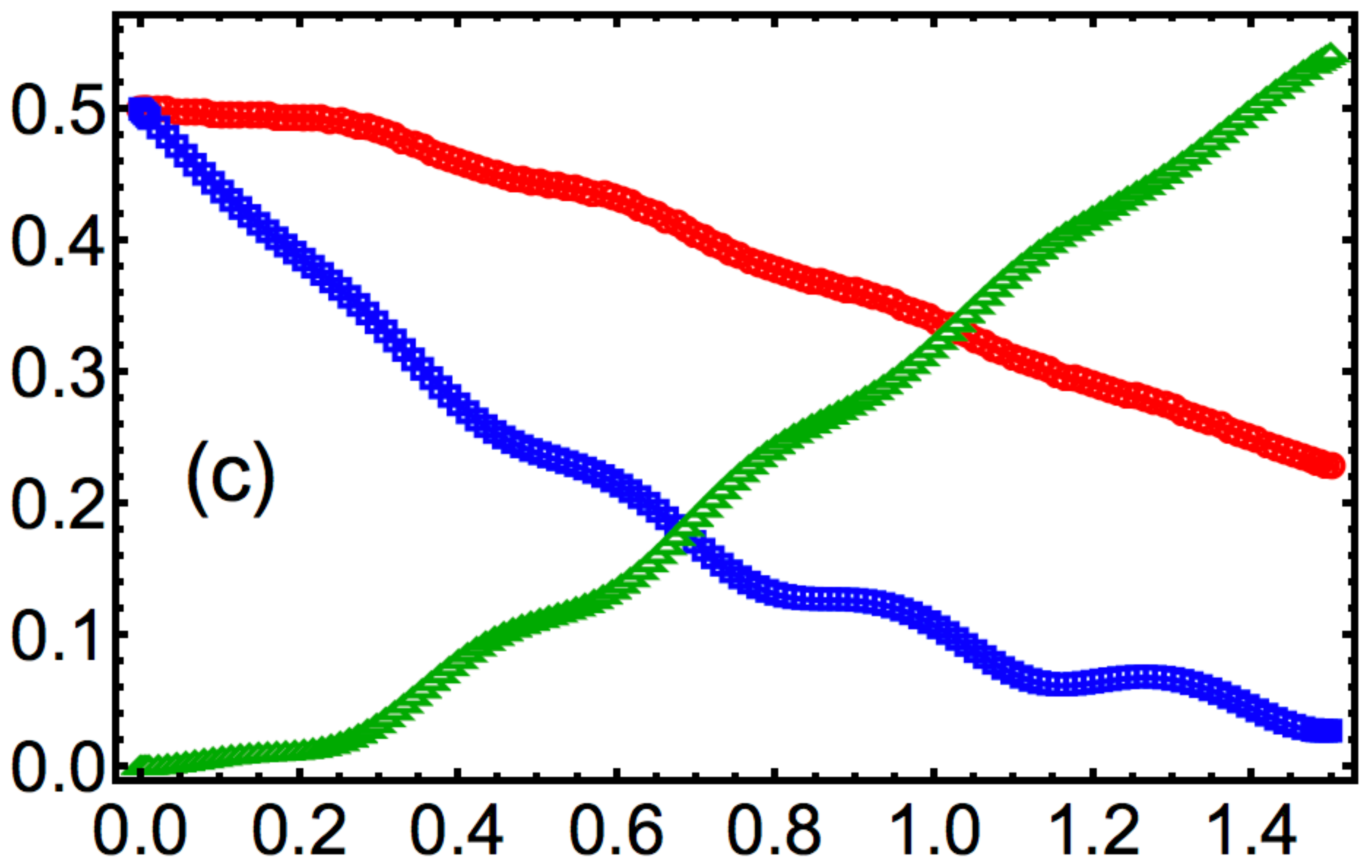}
\includegraphics[width=2.05in, height=1.25in]{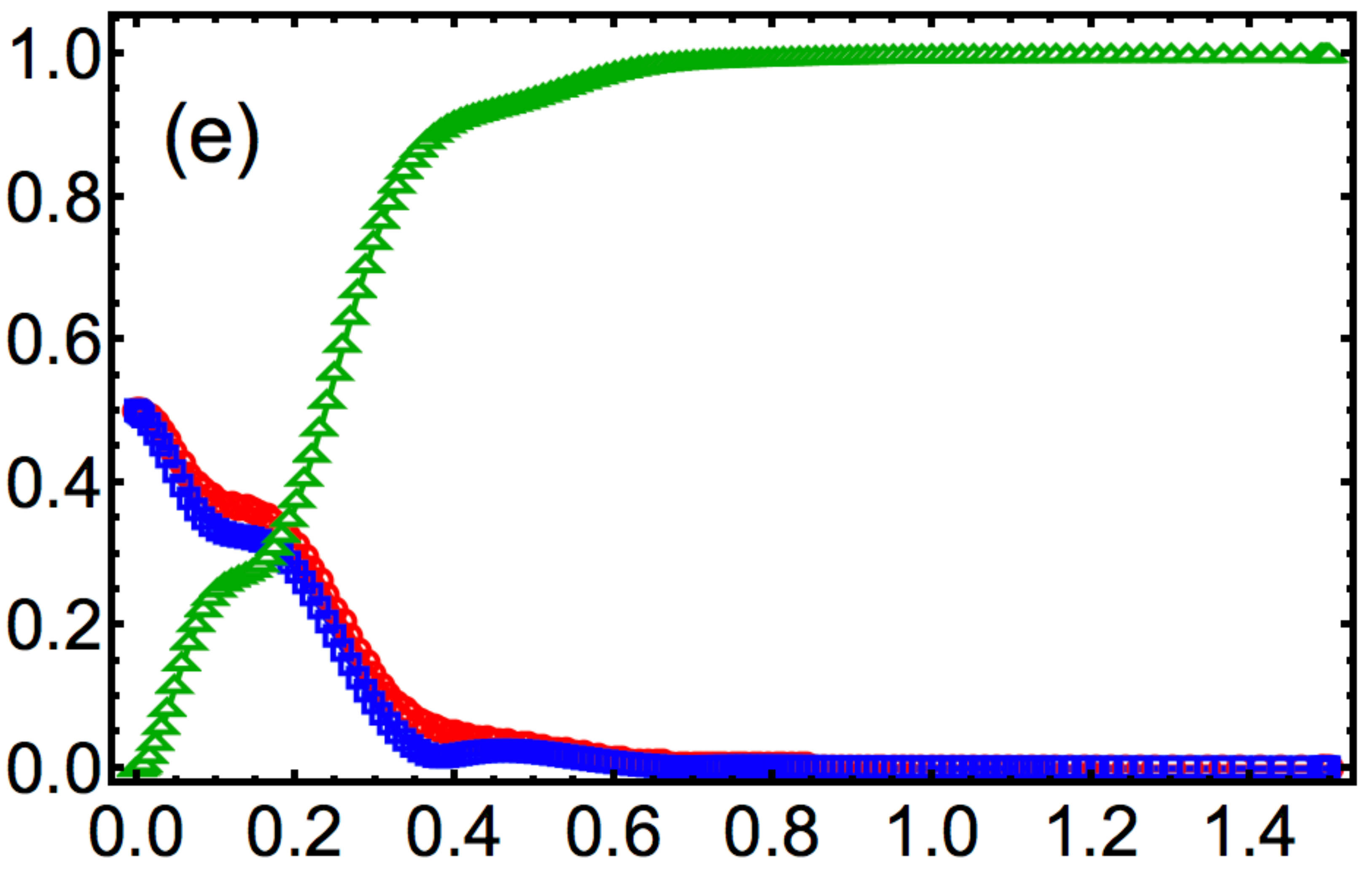}\\
\includegraphics[width=2.05in, height=1.4in]{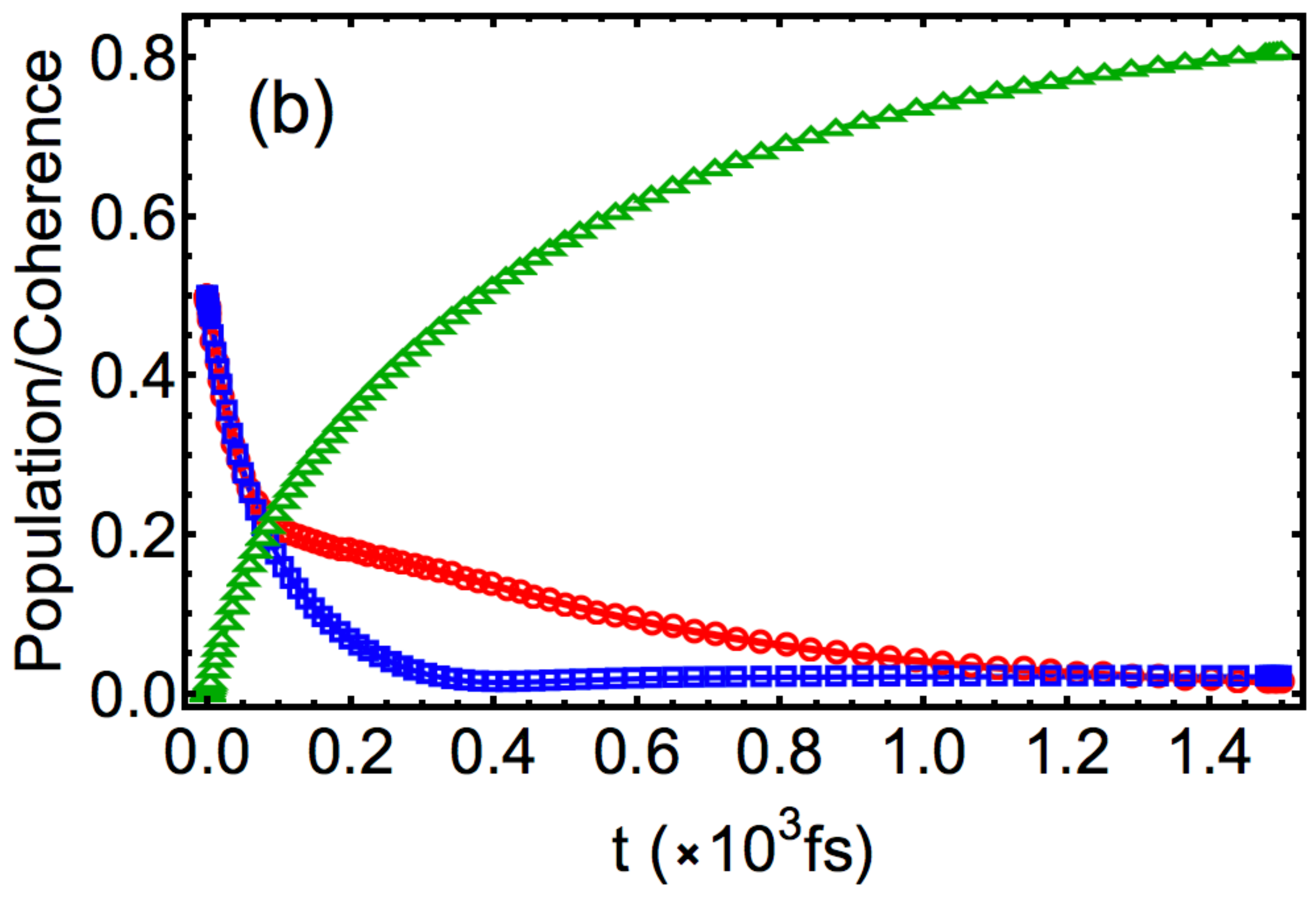}
\includegraphics[width=2.05in, height=1.4in]{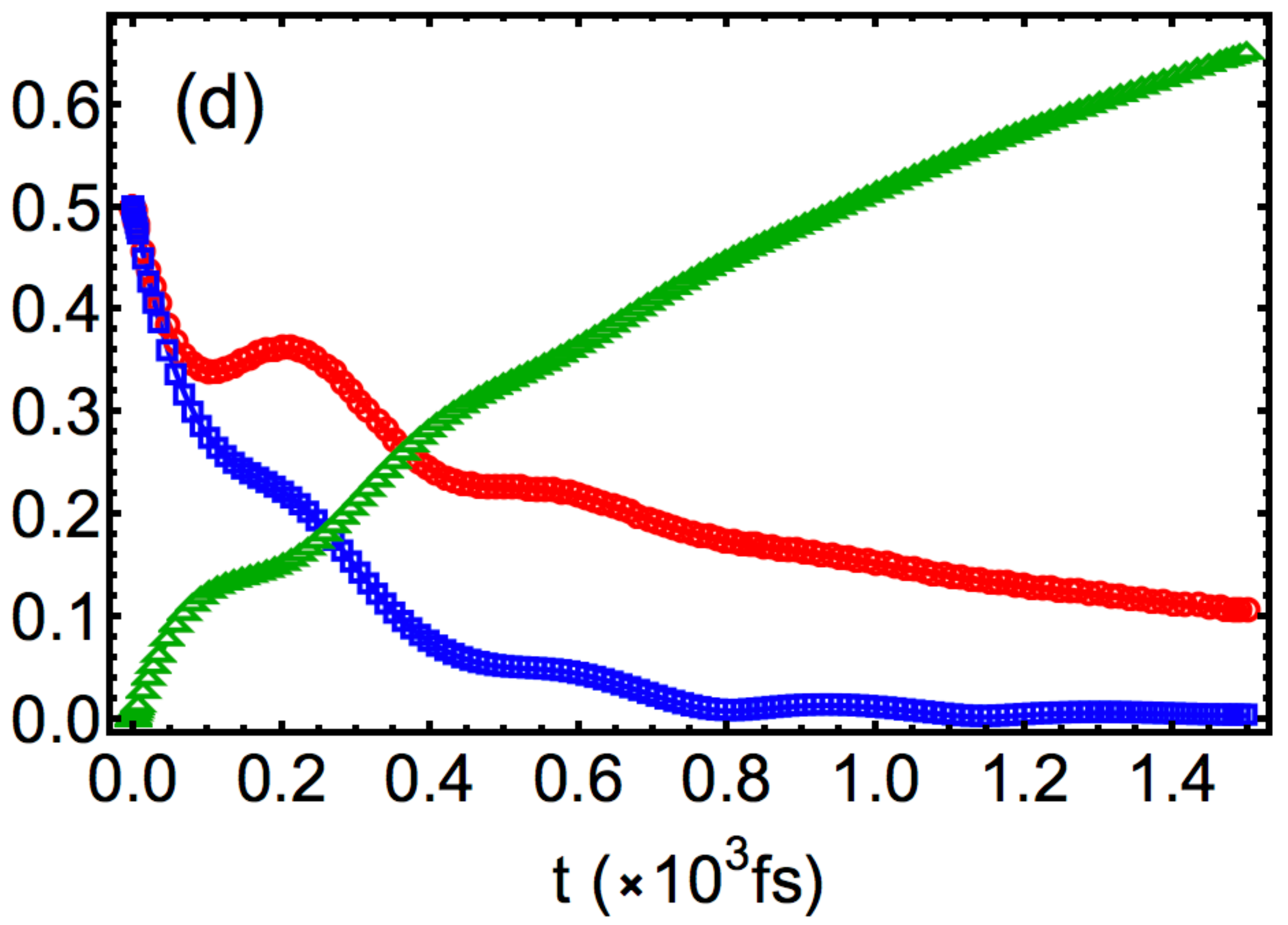}
\includegraphics[width=2.05in, height=1.4in]{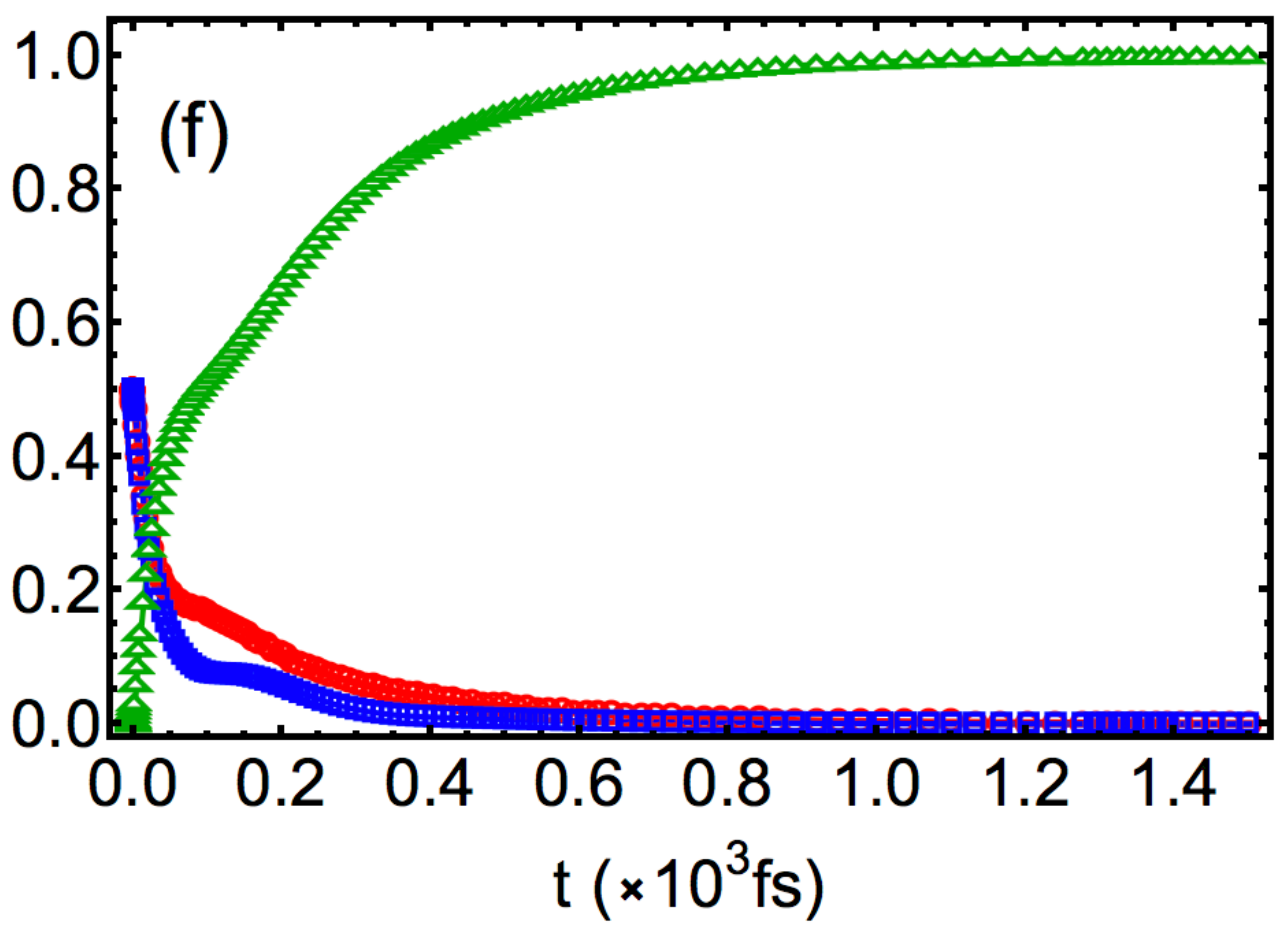}
\captionsetup{
  format=plain,
  margin=1em,
  justification=raggedright,
  singlelinecheck=false
}
\caption{Top (bottom) row shows the population and coherence dynamics in non-Markovian (Markovian) case. Plots pair ((a), (b)), ((c), (d)), and ((e), (f)) represents the overdamped, underdamped, and intermediately damped regimes, respectively.}
\label{fig:pltNpop}
\end{figure*}

In Fig.~\ref{fig:pltNpop} we present the system dynamics in a panel format obtained from the numerical solution of our five-level TCL master equation (\eqref{NMME5}). The relevant time scale of the problem is in femtoseconds (fs). The top (bottom) row of the panel represents the non-Markovian (Markovian) model, while three columns show three parameter regimes (overdamped, underdamped, and intermediate) summarized in Table.~\ref{paratable}. Out of various density matrix elements involved in the problem, we primarily focus on the element $\rho_{e_1e_2}$ which represents the quantum coherence generated between states $\ket{e_1}$ and $\ket{e_2}$ due to the coupling with a common bath via the ground state $\ket{g}$ channel. This so-called noise-induced coherence has been argued to be not a hinder but to aid the higher photosynthetic yield in the Markovian case \cite{dorfman2013photosynthetic}. Here we focus on the behavior of this coherence due to non-Markovian baths and ask the question which type of bath (Markovian or non-Markovian) results in a more efficient excitation energy transfer? As the initial conditions, we suppose 
\begin{align}
\rho_{e_2e_2}=0.5,~\text{and}~\rho_{e_1e_2}=0.5   
\end{align}
with all other density matrix elements being zero in the beginning. From Fig.~\ref{fig:pltNpop}, we first notice that in both Markovian and non-Markovian cases the coherence starts from its initial value, and then gradually decreases as a function of time. Next, we observe two key results: (1) In all three working regimes, there is a considerable enhancement in the coherence magnitude $|\rho_{e_1e_2}|$ as we consider the non-Markovian case. (2) However, in the non-Markovian scenario we also observe the meta-stable state $\ket{m_1}$ population (i.e. $\rho_{m_1m_1}$) exhibit relatively slow build-up compared to its Markovian counterpart. This trend indicates that in the Markovian case even though a larger coherence can be generated, it takes a longer time for the charge separation to grow as compared to the Markovian case.

\begin{figure}[h]
\centering
\includegraphics[width=0.48\textwidth]{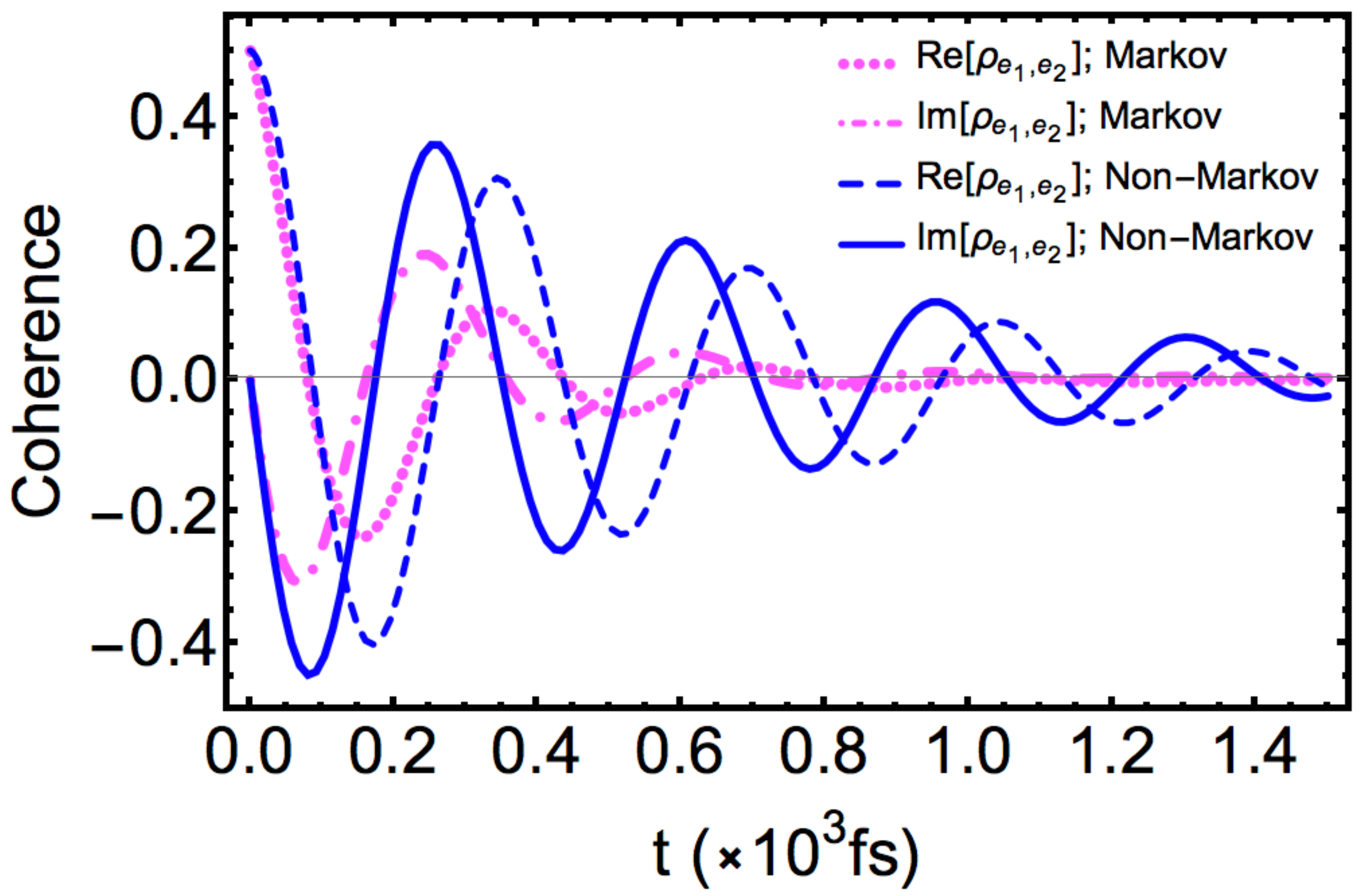}
\captionsetup{
  format=plain,
  margin=1em,
  justification=raggedright,
  singlelinecheck=false
}
\caption{Comparison of the temporal evolution of real and imaginary parts of coherence in the Markovian and non-Markovian cases in the underdamped regime. }
\label{Fig4}
\end{figure}

In Fig.~\ref{Fig4} we report the time evolution of real and imaginary parts of $\rho_{e_1e_2}$. The dotted magenta, dotted-dashed magenta, dashed blue, and solid blue curves represent the Markovian $\mathrm{Re}[\rho_{e_1e_2}]$, Markovian $\mathrm{Im}[\rho_{e_1e_2}]$, non-Markovian $\mathrm{Re}[\rho_{e_1e_2}]$, and non-Markovian $\mathrm{Im}[\rho_{e_1e_2}]$ cases, respectively. Inspired by the Fig.~\ref{fig:pltNpop} (c) and (d), in Fig.~\ref{Fig4} we have selected the underdamped regime which allows the survival of higher coherence in the long-time limit of non-Markovian regime. Both $\mathrm{Re}[\rho_{e_1e_2}](t)$, $\mathrm{Im}[\rho_{e_1e_2}](t)$ curves confirm that the non-Markovian case not only maintains a higher coherence for all transient times, but also reaches a steady-state value (around $1500 fs$) that surpasses the corresponding Markovian case value. This behavior opens the possibility of further improvement in the photosynthetic yield due to non-Markovianity.

\section{Summary and Conclusions}
In summary, by questioning the validity of Markov approximation in the open quantum system treatment of photosynthetic reaction centers in this work, we studied the impact of memory-full environments (non-Markovianity) on the time evolution of noise-induced coherence and population dynamics in type-II photosystems. Therein, we model the photosystem-II as a five-level QHE and studied the dynamics of the photosynthetic process through a local-in-time non-Markovian master equation of the TCL type. As an example, we concentrated on the experimentally feasible Lorentzian spectral-shaped non-Markovian environments that lead to time-dependent transition rates. As the main finding of our calculations, we concluded that the Lorentzian environments can enhance the noise-induced coherence (in all working regimes under transient conditions and the underdamped regime under long-time situations) as compared to the corresponding Markovian case. This result can improve our understanding of nearly perfect efficiency attained by nature in the photosynthesis processes. Additionally, this work may find applications in engineering next-generation efficient quantum devices that are bio-inspired.\\ 

\hspace{-5mm}{\bf ACKNOWLEEDGMENTS} \\
IMM would like to acknowledge the support of the Miami University College of Arts \& Science and Physics Department start-up funding.

\bibliographystyle{ieeetr}
\bibliography{paper}
\end{document}